\documentclass[english,twocolumn, nofootinbib,citeautoscript]{revtex4-2}
\usepackage[T1]{fontenc}
\usepackage[latin9]{inputenc}
\usepackage{color}
\usepackage{babel}
\usepackage{amsmath}
\usepackage{amssymb}
\usepackage{graphicx}
\usepackage[unicode=true,pdfusetitle,
 bookmarks=true,bookmarksnumbered=false,bookmarksopen=false,
 breaklinks=false,pdfborder={0 0 0},pdfborderstyle={},backref=false,colorlinks=true]
 {hyperref}

\makeatletter
\newcommand{\argmax}{\mathop{\mathrm{argmax}}}

\usepackage{mhchem}

\renewcommand{\fnum@figure}{\footnotesize {\bf Figure \thefigure}}

\makeatletter
\g@addto@macro\bfseries{\boldmath}
\makeatother

\RequirePackage{xcolor} %
\definecolor{darkgreen}{rgb}{0, 0.4, 0} 
\definecolor{midgreen}{rgb}{0.5, 0.8, 0.5}
\definecolor{darkred}{rgb}{0.6, 0, 0}
\definecolor{darkblue}{rgb}{0, 0, 0.6} 

\hypersetup{linktocpage=true, colorlinks=true, citecolor={darkgreen}, linkcolor={darkred}, urlcolor={darkblue} }

\makeatother

\begin{document}
\title{{\Large Far from Asymptopia:}\vspace{1mm} \\
Unbiased High-Dimensional Inference Cannot Assume Unlimited Data}
\author{Michael C. Abbott}
\affiliation{Department of Physics, Yale University, New Haven CT}
\author{Benjamin B. Machta\vspace{1mm}}
\affiliation{Department of Physics, Yale University, New Haven CT}
\date{v2: 27 February 2023}
\begin{abstract}
\noindent Inference from limited data requires a notion of measure
on parameter space, most explicit in the Bayesian framework as a prior.
Here we demonstrate that Jeffreys prior, the best-known uninformative
choice, introduces enormous bias when applied to typical scientific
models. Such models have a relevant effective dimensionality much
smaller than the number of microscopic parameters. Because Jeffreys
prior treats all microscopic parameters equally, it is from uniform
when projected onto the sub-space of relevant parameters, due to variations
in the local co-volume of irrelevant directions. We present results
on a principled choice of measure which avoids this issue, leading
to unbiased inference in complex models. This optimal prior depends
on the quantity of data to be gathered, and approaches Jeffreys prior
in the asymptotic limit. However, this limit cannot be justified without
an impossibly large amount of data, exponential in the number of microscopic
parameters. 
\end{abstract}
\maketitle

\section*{Introduction}

\noindent No experiment fixes a model's parameters perfectly. Every
approach to propagating the resulting uncertainty must, explicitly
or implicitly, assume a measure on the space of possible parameter
values. A badly chosen measure can introduce bias, and we argue here
that avoiding such bias is equivalent to the very natural goal of
assigning equal weight to each distinguishable outcome. However, this
goal is seldom reached, either because no attempt is made, or because
the problem is simplified by prematurely assuming the asymptotic limit
of nearly infinite data. We demonstrate here that this assumption
can lead to large bias in what we infer from the parameters, in models
with features typical of many-parameter mechanistic models found in
science. We propose a score for such bias, and advocate for using
a measure which makes this zero. Such a measure allows for unbiased
inference without the need to first simplify the model to just the
right degree of complexity. Instead, weight is automatically spread
according to a lower effective dimensionality, ignoring details irrelevant
to visible outcomes.

We consider models which predict a probability distribution $p(x|\theta)$
for observing data $x$ given parameters $\theta$. The degree of
overlap between two such distributions indicates how difficult it
is to distinguish the two parameter points, which gives a notion of
distance on parameter space. The simplifying idea of information geometry
is to focus on infinitesimally close parameter points, for which there
is a natural Riemannian metric, the Fisher information \citep{rao1945information,amari1983foundation}.
This may be thought of as having units of standard deviations, so
that along a line of integrated length $L$ there are about $L$ distinguishable
points, and thus any parameter which can be measured to a few digits
precision has length $L>100$. It is a striking empirical feature
of models in science that most have a few such long (or relevant)
parameter directions, followed by many more short (or irrelevant)
orthogonal directions \citep{brown2003statistical,daniels2008sloppiness,machta2013parameter,quinn2023information}.
The irrelevant lengths, all $L<1$, show a characteristic spectrum
of being roughly evenly spaced on a log scale, often over many decades.
As a result, much of the geometry of this Riemannian model manifold
consists of features much smaller than 1, far too small to observe.
But the natural intrinsic volume measure which follows from the Fisher
metric is sensitive to all of these unobservable dimensions, and as
we demonstrate here, they cause this measure to introduce enormous
bias.

\begin{figure}
\includegraphics[width=1\columnwidth]{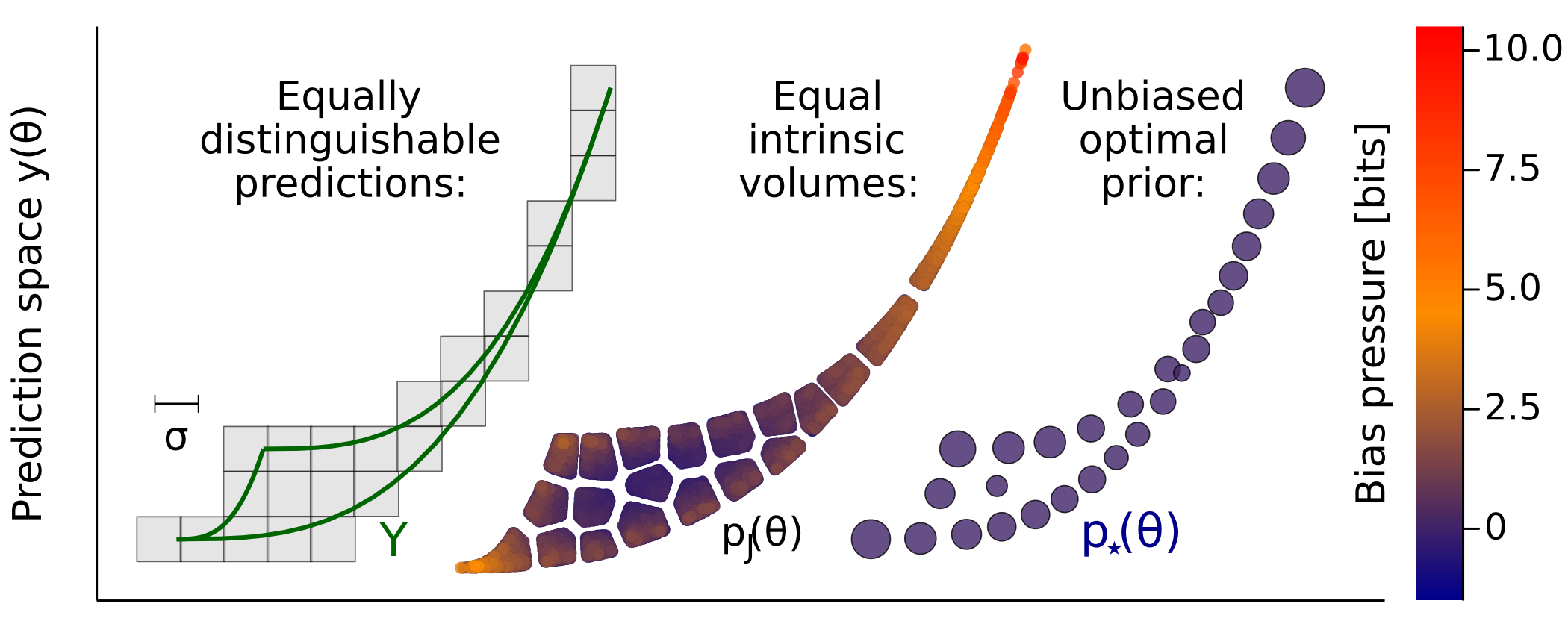}

\caption{The natural volume is a biased measure for the space of distinguishable
outcomes. The left panel outlines the space of possible predictions
$Y$; the observed $x$ is deterministic $y(\theta)$ plus measurement
noise. With the scale of the noise $\sigma$ as shown, the upper half
is effectively one-dimensional. The centre panel shows a sample from
the volume measure $p_{\mathrm{J}}(\theta)$, divided into blocks
of equal weight. These are strongly influenced by the unobservable
thickness of the upper portion. Points are coloured by bias pressure
$b(\theta)$ which we define in equation (\ref{eq:defn-bias}). The
right panel shows the explicitly unbiased optimal measure $p_{\star}(\theta)$,
which gradually adjusts from two- to one-dimensional behaviour.\label{fig:summary}
(The model is equation (\ref{eq:defn-exp-model}) with $a_{1}=0.8$,
$a_{2}=0.2$, and $k_{1}\protect\geq k_{2}$, observed at times $t=1,3$
each with Gaussian noise $\sigma=0.1$.)}
\end{figure}

To avoid this problem, we need a measure tied to Fisher length scale
$L\approx1$, instead of one from the continuum. Locally, this length
scale partitions dimensions into relevant and irrelevant, which in
turn approximately factorises the volume element into a relevant part
and what we term the \emph{irrelevant co-volume}. The wild variations
of this co-volume are the source of the bias we describe, and it is
rational to ignore them.  As we illustrate in figure \ref{fig:summary}
for a simple two-parameter model, equally distinguishable predictions
do not correspond to equal intrinsic volumes, and this failure is
detected by a score we call \emph{bias pressure}. The measure $p_{\star}(\theta)$
for which this score is everywhere zero, by contrast, captures relevant
distinguishability and ignores the very thin irrelevant direction.
The same measure is also obtained by maximising the information learned
about parameter $\theta$ from seeing data $x$ \citep{lindley1961use,bernardo1979reference,mattingly2018maximizing},
or equivalently from a particular minimax game \citep{kashyap1971prior,haussler1997general,krob1997minimax}.
Since $p_{\star}(\theta)$ is usually discrete \citep{Farber:1967us,Smith:1971kt,berger1989priors,Zhang:1994ui,scholl1998shannon,sims2006rational,mcdonnell2009information,mattingly2018maximizing},
it can be seen as implementing a length cutoff, replacing the smooth
differential-geometric view of the model manifold with something quantised
\citep{connes1994noncommutative}.

In the Bayesian framework, the natural continuous volume measure $p_{\mathrm{J}}(\theta)$
is known as Jeffreys prior, and is the canonical example of an uninformative
prior: a principled, ostensibly neutral choice. It was first derived
based on invariance considerations \citep{jeffreys1946invariant},
and can also be justified by information- or game-theoretic ideas,
provided these are applied in the limit of infinitely many repetitions
\citep{lindley1961use,bernardo1979reference,clarke1994jeffreys,scholl1998shannon,balasubramanian1997statistical}.
 This asymptotic limit often looks like a technical trick to simplify
derivations. However, in realistic models this limit is very far from
being justified, exponentially far in the number of parameters, often
requiring an experiment to be repeated for longer than the age of
the universe. We demonstrate here that using the prior derived in
this limit introduces large bias, in such models. And we argue that
such bias, and not only computational difficulties, has prevented
the wide use of uninformative priors.

The promise of principled ways of tracking uncertainty, Bayesian or
otherwise, is to free us from the need to select a model with precisely
the right degree of complexity. This idea is often encountered in
the context of overfitting, where the maximum likelihood point of
an overly complex model gives worse predictions. The bias discussed
here is a distinct way for overly complex models to give bad predictions.
We begin with toy models in which the number of parameters can be
easily adjusted. But in the real models of interest, we cannot trivially
tune the number of parameters. This is why we wish to find principled
methods which are not fooled by the presence of many irrelevant parameters.

\section*{Results}

\begin{figure*}
\includegraphics[width=1\textwidth]{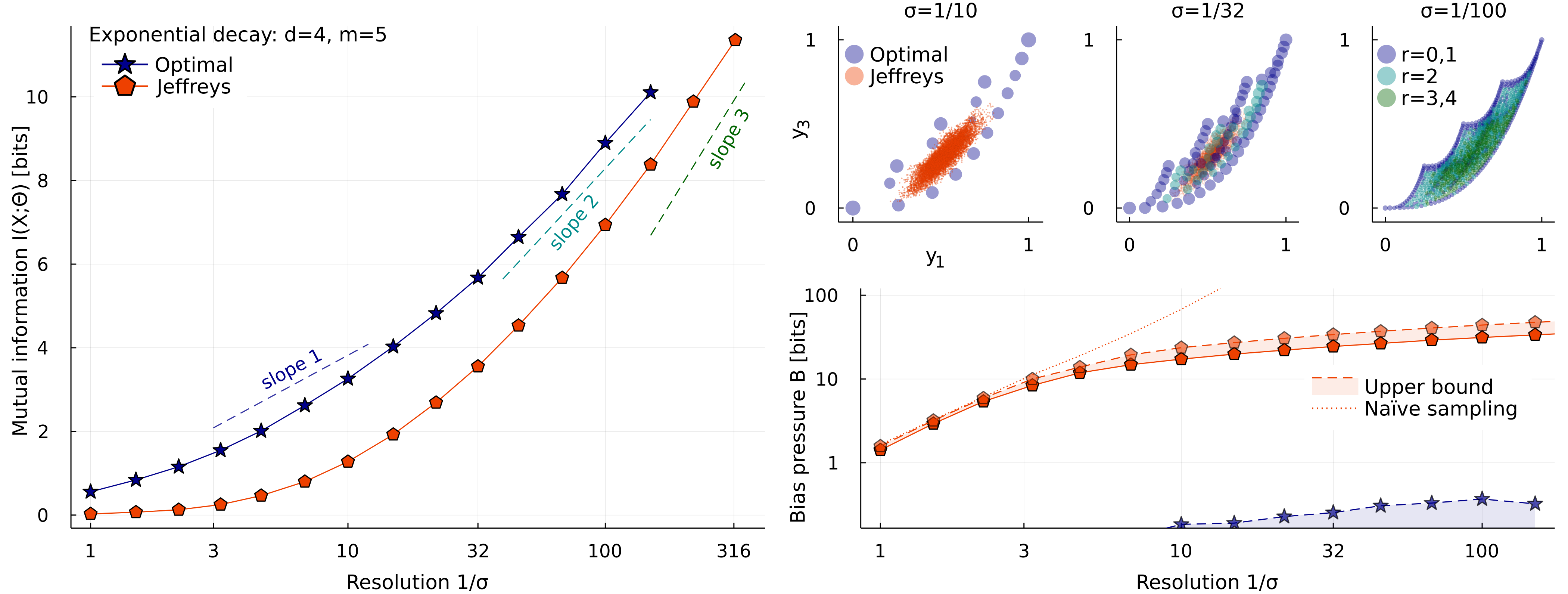}

\caption{The effect of varying the noise level $\sigma$ on a fixed model.
The model is (\ref{eq:defn-exp-model}) with $d=4$ parameters, observed
at $m=5$ times $t=1,2,\ldots5$. Top right, the optimal prior $p_{\star}(\theta)$
has all of its weight on 0- and 1-dimensional edges at large $\sigma$,
but adjusts to fill in the bulk at small $\sigma$. (Colours indicate
the dimension $r$ of the 4-dimensional shape's edge on which a point
is located, the rank of the FIM there.) Jeffreys prior $p_{\mathrm{J}}(\theta)$
is independent of $\sigma$, and has nonzero density everywhere, but
a sample of $10^{6}$ points is largely located near the middle of
the shape. Left, the slope of $I_{\star}(X;\Theta)\sim d_{\mathrm{eff}}\log1/\sigma$
gives a notion of effective dimensionality; in the asymptotic limit
$\sigma\to0$ we expect $d_{\mathrm{eff}}=d=4$. Bottom right, the
worst-case bias pressure $B=\max_{\theta}b(\theta)$ is always zero
for $p_{\star}(\theta)$, up to numerical error, but remains nonzero
for $p_{\mathrm{J}}(\theta)$ even in the asymptotic limit. The appendix
describes how upper and lower bounds for $B$ are calculated.\label{fig:vary-sigma}}
\end{figure*}

\noindent We consider a model to be characterised by the likelihood
$p(x|\theta)$ of observing data $x\in X$ when the parameters are
$\theta\in\Theta$. In such a model, the Fisher information metric
(FIM) measures distinguishability of nearby points in parameter space
as a distance $ds^{2}(\theta)=\sum_{\mu,\nu=1}^{d}g_{\mu\nu}(\theta)d\theta^{\mu}d\theta^{\nu}$,
where
\begin{equation}
g_{\mu\nu}(\theta)=-\int\negmedspace dx\:p(x|\theta)\:\partial_{\mu}\partial_{\nu}\log p(x|\theta).\label{eq:defn-FIM}
\end{equation}
For definiteness we may take points separated along a geodesic by
a distance $L=\int\sqrt{ds^{2}(\theta)}>1$ to be distinguishable.
Intuitively, though incorrectly, the $d$-dimensional volume implied
by the FIM might be thought to correspond to the total number of distinguishable
parameter values inferable from an experiment:
\[
Z=\int\negmedspace d\theta\sqrt{\det g(\theta)}.
\]
However, this counting makes a subtle assumption, that all structure
in the model has a scale much larger than 1. When many dimensions
are smaller than 1, their lengths weight the effective volume along
the larger dimensions, despite having no influence on distinguishability.

The same effect applies to the normalised measure, Jeffreys prior:
\begin{equation}
p_{\mathrm{J}}(\theta)=\frac{1}{Z}\sqrt{\det g(\theta)}.\label{eq:defn-Jeff}
\end{equation}
This measure's dependence on the \emph{irrelevant co-volume} is an
under-appreciated source of bias in posteriors derived from this prior.
The effect is most cleanly seen when the FIM is block-diagonal, $g=g_{\mathrm{rel}}\oplus g_{\mathrm{irrel}}$.
Then the volume form factorises exactly, and the relevant effective
measure is the $\sqrt{\det g_{\mathrm{rel}}(\theta_{\mathrm{rel}})}$
factor times $V_{\mathrm{irrel}}(\theta_{\mathrm{rel}})$, an integral
over the irrelevant dimensions.

A more principled measure of the (log of the) number of distinguishable
outcomes is the mutual information between parameters and data, $I(X;\Theta)$:
\[
I(X;\Theta)=\int\negmedspace d\theta\:p(\theta)D_{\mathrm{KL}}\big[p(x|\theta)\big\Vert p(x)\big]
\]
where $D_{\mathrm{KL}}$ is the Kullback--Leibler divergence between
two probability distributions, which are not necessarily close: $p(x)=\int\negmedspace d\theta\:p(\theta)\:p(x|\theta)$
is typically much broader than $p(x|\theta)$. Unlike the volume $Z$,
the mutual information depends on the prior $p(\theta)$. Past work
both by ourselves and others has advocated for using the prior which
maximizes this mutual information, with \citep{bernardo1979reference}
or without \citep{mattingly2018maximizing} taking the asymptotic
limit:
\begin{equation}
p_{\star}(\theta)=\mathop{\mathrm{argmax}}_{p(\theta)}I(X;\Theta).\label{eq:pstar-max-MI}
\end{equation}
The same prior arises from a minimax game in which you choose a prior,
your Opponent chooses the true $\theta$, and you lose the (large)
KL divergence \citep{kashyap1971prior,haussler1997general,krob1997minimax}:
\begin{equation}
p_{\star}(\theta)=\mathop{\mathrm{argmin}}_{p(\theta)}\max_{\theta}D_{\mathrm{KL}}\big[p(x|\theta)\big\Vert p(x)\big].\label{eq:pstar-game}
\end{equation}
Here we stress a third perspective, defining a quantity we call \emph{bias
pressure} which captures how strongly the prior disfavours predictions
from a given point:
\begin{equation}
b(\theta)=\frac{\partial I(X;\Theta)}{\partial p(\theta)}\Big\vert_{\int\negmedspace d\theta\,p(\theta)=1}=D_{\mathrm{KL}}\big[p(x|\theta)\big\Vert p(x)\big]-I(X;\Theta).\label{eq:defn-bias}
\end{equation}
The optimal $p_{\star}(\theta)$ has $b(\theta)=0$ on its support,
and can be found by minimising $B=\max_{\theta}b(\theta).$ Other
priors have $b(\theta)>0$ at some points, indicating that $I(X;\Theta)$
can be increased by moving weight there (and away from points where
$b(\theta)<0$). We demonstrate below that $b(\theta)$ deserves to
be called a bias, as it relates to large deviations of the posterior
centre of mass. We do this by presenting a number of toy models, chosen
to have information geometry similar to that typically found in mechanistic
models from many scientific fields \citep{machta2013parameter}.

\subsection*{Exponential decay models}

\noindent The first model we study involves inferring rates of exponential
decay. This may be motivated for instance by the problem of determining
the composition of a radioactive source containing elements with different
half-lives, using Geiger counter readings taken over some period of
time. The mean count rate at time $t$ is 
\begin{equation}
y_{t}(\theta)=\sum_{\mu=1}^{d}a_{\mu}e^{-k_{\mu}t},\quad k_{\mu}=e^{-\theta_{\mu}}>0.\label{eq:defn-exp-model}
\end{equation}
We take the decay rates as parameters, and fix the proportions $a_{\mu}$,
usually to $a_{\mu}=1/d$, thus initial condition $y_{0}=1$. If we
make observations at $m$ distinct times $t$, then the prediction
$y$ is an $m$-vector, restricted to a compact region $Y\subset[0,1]^{m}$.
For radioactivity we would expect to observe $y_{t}$ plus Poisson
noise, but the qualitative features are the same if we simplify to
Gaussian noise with constant width $\sigma$:
\begin{equation}
p(x|\theta)=e^{-\left|x-y(\theta)\right|^{2}/2\sigma^{2}}\big/(2\pi\sigma^{2})^{m/2}.\label{eq:defn-gauss-like}
\end{equation}
The Fisher metric then simplifies to be the Euclidean metric in the
space of predictions $Y$, pulled back to parameter space $\Theta$:
\[
g_{\mu\nu}(\theta)=\frac{1}{\sigma^{2}}\sum_{t,t'}^{m}\frac{\partial y_{t}}{\partial\theta^{\mu}}\frac{\partial y_{t'}}{\partial\theta^{\nu}}\:\delta_{tt'}
\]
thus plots of $p(y)$ in $\mathbb{R}^{m}$ will show Fisher distances
accurately. This model is known to be ill-conditioned, with many small
manifold widths, and many small FIM eigenvalues, when $d$ is large
\citep{transtrum2010why}.

\begin{figure*}
\includegraphics[width=1\textwidth]{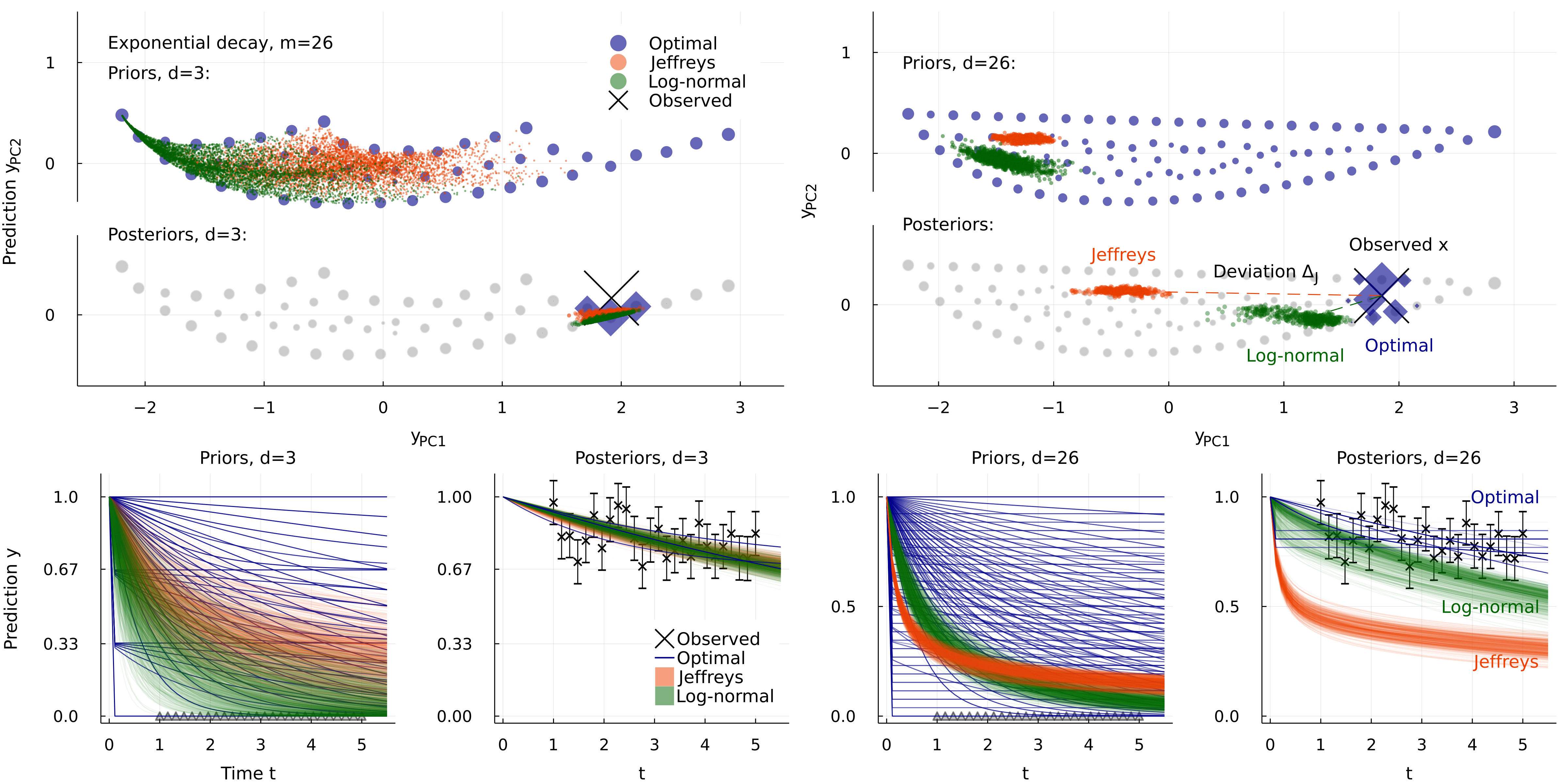}

\caption{The effect of changing model dimension, for fixed data and noise level.
Left half, the exponential decay model of equation (\ref{eq:defn-exp-model})
with $d=3$ parameters, observed with noise $\sigma=0.1$ at $m=26$
times in $1\protect\leq t\protect\leq5$. Three priors are shown,
drawn above by projecting onto the first two principal components
of vector $y$, and below-left as a time-course $y_{t}$. (Each point
on the upper plot is a line on the lower one.) The corresponding posteriors
shown for a particular fixed $x$, which is the large cross in the
upper plot (where the prior is shown again in light gray as a visual
guide) and the series of points in the lower plot. In the $d=3$ model,
all three posteriors are reasonable fits to the data. Right half,
the similar model with $d=26$ parameters, for the same observations
with the same noise. Here Jeffreys prior is much more strongly concentrated,
favouring the part of the manifold where the irrelevant dimensions
are largest. This has the effect of biasing the posterior far from
the data, more than 20 standard deviations away. Figures \ref{fig:posterior-deviation}
and \ref{fig:scores-vs-d} explore the same setup further, including
intermediate dimensions $d$. The log-normal prior is introduced in
equation (\ref{eq:log-normal-prior}). \label{fig:posterior-3-26}}
\end{figure*}

With just two dimensions, $d=m=2$, figure \ref{fig:summary} shows
the region $Y\subset\mathbb{R}^{2}$, Jeffreys prior $p_{\mathrm{J}}(\theta)$
and the optimal prior $p_{\star}(\theta)$, projected to densities
on $Y$. Jeffreys is uniform $p_{\mathrm{J}}(y)\propto1$ (since the
metric is constant in $y$), and hence always weights a two-dimensional
area, both where this is appropriate and where it's not. The upper
portion of $Y$ in the figure is thin compared to $\sigma$, so the
points we can distinguish are those separated vertically: the model
is effectively one-dimension there. Jeffreys does not handle this
well, which we illustrate in two ways. First, the prior is drawn divided
into 20 segments of equal weight (equal area), which roughly correspond
to distinguishable differences where the model is two-dimensional,
but not where it becomes one-dimensional. Second, the points are coloured
by $b(\theta)$, which detects this effect, and gives large values
at the top (about 10 bits). The optimal prior avoids these flaws,
by smoothly adjusting from the one- to the two-dimensional part of
the model \citep{mattingly2018maximizing}.

The claim that some parts of the model are effectively one-dimensional
depends on the amount of data gathered. Independent repetitions of
the experiment have overall likelihood $p(x^{M}|\theta)=\prod_{i=1}^{M}p(x^{(i)}|\theta)$,
which will always scale the FIM   by $M$, hence all distances by
a factor $\sqrt{M}$. This scaling is exactly equivalent to smaller
Gaussian noise $\sigma$. Increasing $M$ increases the number of
distinguishable points, and large enough $M$ (or small enough $\sigma$)
can eventually make any nonzero length larger than 1. Thus the amount
of data gathered affects which parameters are relevant. But notice
that such repetition has no effect at all on $p_{\mathrm{J}}(\theta)$,
since the scale of $g_{\mu\nu}(\theta)$ in equation (\ref{eq:defn-Jeff})
is cancelled by $Z$. In this sense it is already clear that Jeffreys
prior belongs to the fixed point of repetition, i.e. to the asymptotic
limit $M\to\infty$.

Figure \ref{fig:vary-sigma} shows a more complicated version of the
model (\ref{eq:defn-exp-model}), with $d=4$ parameters, and looks
at the effect of varying the noise level $\sigma$. Jeffreys prior
always fills the 4-dimensional bulk, but at moderate $\sigma$, most
of the distinguishable outcomes are located far from this mass. 
At large $\sigma$, equivalent to few repetitions, all the weight
of the optimal prior is on zero- and one-dimensional edges. As more
data is gathered, it gradually fills in the bulk, until in the asymptotic
limit $\sigma\to0$, it approaches Jeffreys prior \citep{clarke1990informationtheoretic,clarke1994jeffreys,krob1997minimax,scholl1998shannon}.
However, while $p_{\star}(\theta)$ approaches a continuum at any
interior point \citep{abbott2019scaling}, it remains discrete at
Fisher distances $\sim1$ from the boundary.The worst-case bias pressure
detects this, hence the maximum for Jeffreys prior does not approach
that for the optimal prior: $B_{\mathrm{J}}\not\to0$. However, since
mutual information is dominated by the interior in this limit, we
expect the values for $p_{J}(\theta)$ and $p_{\star}(\theta)$ to
agree in the limit: $I_{\mathrm{J}}-I_{\star}\to0$.

\begin{figure}
\includegraphics[width=1\columnwidth]{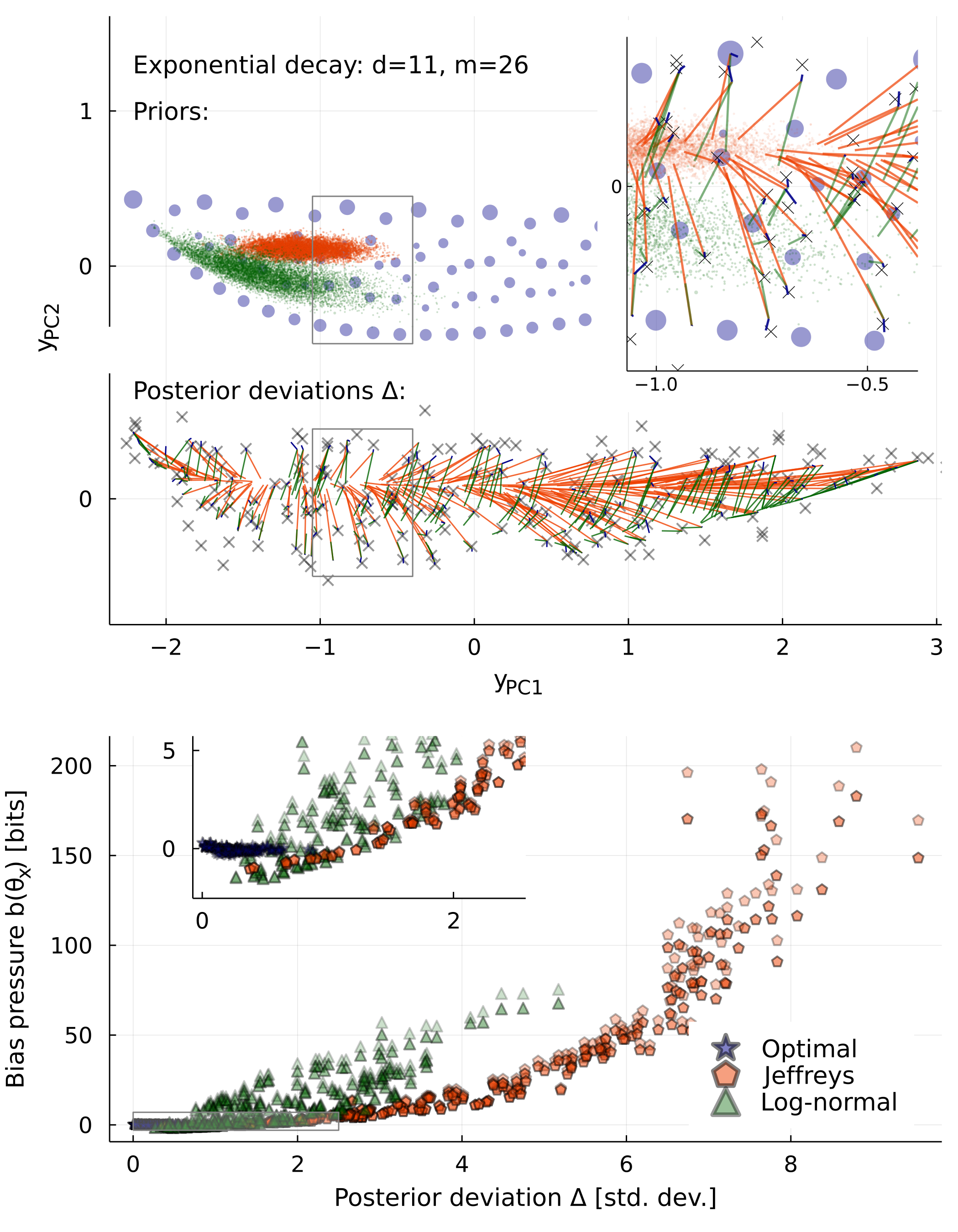}

\caption{Posterior bias due to concentration of measure. Top, priors for the
$d=11$ case of the model in figure \ref{fig:posterior-3-26}. We
calculate the posterior for each at 100 points $x$ (marked), and
draw a line connecting the maximum likelihood point $y(\hat{\theta}_{x})$
to the posterior centre of mass $\left\langle y(\theta)\right\rangle _{x}$.
Inset enlarges to show that there are blue lines too, for the optimal
prior, most much shorter than the spacing of its atoms. Below, we
compare the length of such lines (divided by $\sigma=0.1$) to the
bias pressure $b(\hat{\theta}_{x})$. Notice that $b(\theta)$ is
sometimes negative (it has zero expectation value: $\int d\theta\:p(\theta)\:b(\theta)=0$),
although the worst-case $B=\max_{\theta}b(\theta)$ is non-negative.
Each pair of darker and lighter points are a lower and an upper bound,
explained in the appendix. \label{fig:posterior-deviation}}
\end{figure}

One way to quantify the effective dimensionality is to look at the
rate of increase of mutual information under repetition, or decreasing
noise $\sigma$. Along a dimension with Fisher length $L\gg1$, the
number of distinguishable points is proportional to $L$, and thus
a cube with $d_{\mathrm{eff}}$ large dimensions will have $\propto L^{d_{\mathrm{eff}}}$
such points. This motivates defining $d_{\mathrm{eff}}$ by: 
\begin{equation}
I_{\star}(X;\Theta)\sim d_{\mathrm{eff}}\log L,\qquad L=\int\negmedspace\sqrt{ds^{2}}\propto1/\sigma.\label{eq:d_eff-scaling}
\end{equation}
Figure \ref{fig:vary-sigma} shows lines for slope $d_{\mathrm{eff}}=1,2,3$,
and we expect $d_{\mathrm{eff}}\to d$ in the limit $\sigma\to0$.

\subsection*{The costs of high dimensionality}

\begin{figure}
\includegraphics[width=1\columnwidth]{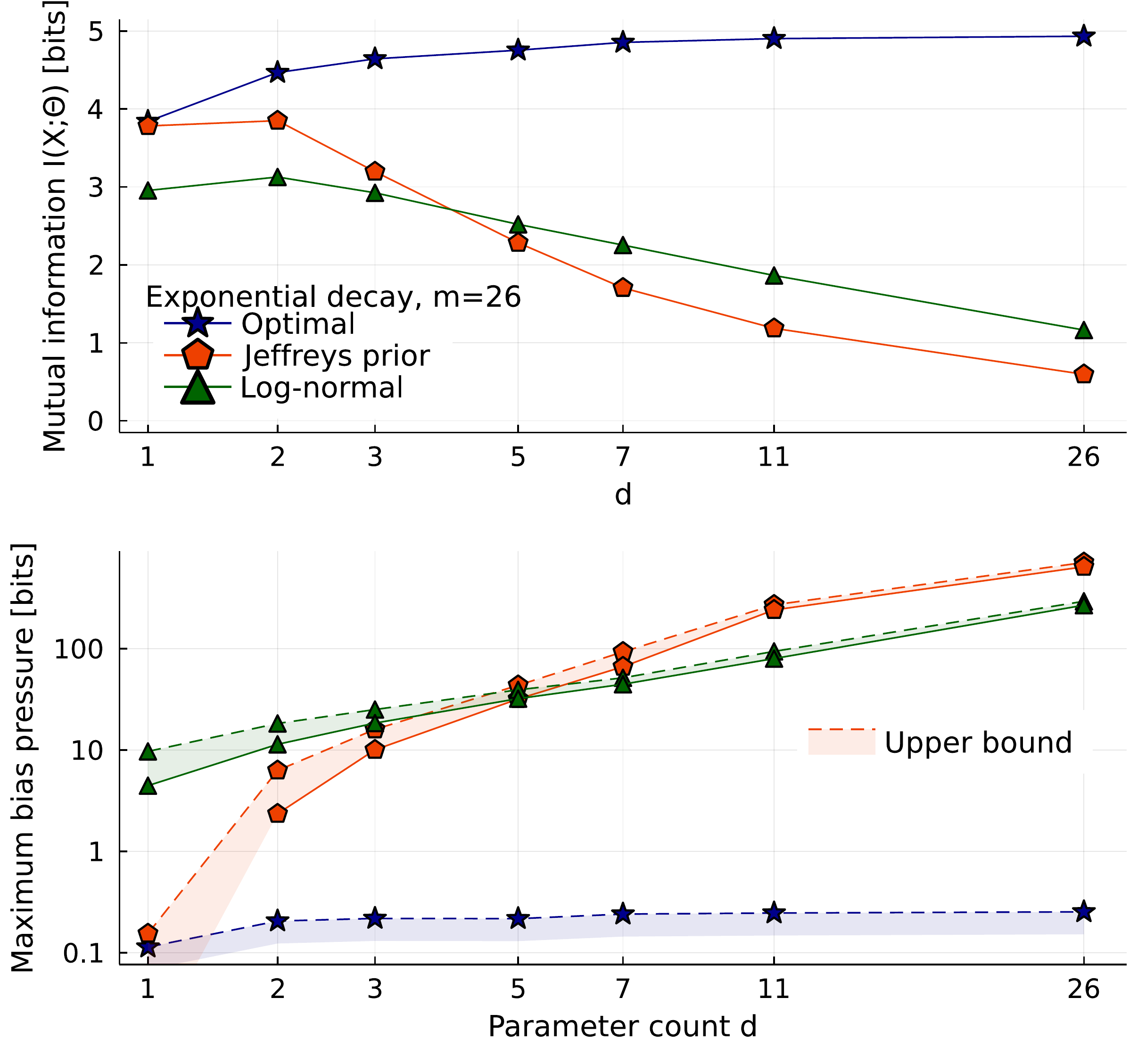}\caption{Information-theoretic scores for priors, as a function of dimensionality
$d$. Like figure \ref{fig:posterior-3-26} these models all describe
the same data, with the same noise. Above, mutual information $I(X;\Theta)/\log2$
(all plots are scaled thus to have units of bits). The optimal prior
ignores the addition of more irrelevant parameters, but Jeffreys prior
is badly affected, and ends up capturing less than 1 bit. Below, worst-case
bias pressure $\max_{\theta}b(\theta)/\log2$. This should be zero
for the optimal prior, but our numerical solution has small errors.
For the other priors, we plot lower and upper bounds, calculated using
Bennett's method \citep{bennett1976efficient}, as described in the
appendix. The bias of Jeffreys prior increases strongly with the increasing
concentration of its weight in higher dimensions.\label{fig:scores-vs-d}}
\end{figure}

\noindent The problems of uneven measure grow more severe with more
dimensions. To explore this, figures \ref{fig:posterior-3-26} to
\ref{fig:scores-vs-d} show a sequence of models with 1 to 26 parameters.
All describe the same data: observations at the same list of $m=26$
times in $1\leq t\leq5$ with the same noise $\sigma=0.1$. While
Jeffreys prior is nonzero everywhere, its weight is concentrated where
the many irrelevant dimensions are largest. With a Monte Carlo sample
of a million points, all are found within the a small orange area
on the right of figure \ref{fig:posterior-3-26}. For a particular
observation $x$, we plot also the posterior $p(\theta|x)$ for each
prior. The extreme concentration of weight in $p_{\mathrm{J}}(\theta)$
in $d=26$ pulls this some 20 standard deviations away from the maximum
likelihood point $y(\hat{\theta}_{x})$. We call this distance the
posterior deviation $\Delta$; it is the most literal kind of bias
in results.

Figure \ref{fig:posterior-deviation} compares the posterior deviation
$\Delta$ to the bias pressure $b(\theta)$ defined in equation (\ref{eq:defn-bias}).
For each of many observations $x$, we find the maximum likelihood
point $\hat{\theta}_{x}=\argmax_{\theta}p(x|\theta)$, and calculate
the distance from this point to the posterior expectation value of
$y$:
\begin{equation}
\Delta(x)=\frac{1}{\sigma}\Big|y(\hat{\theta}_{x})-\int\negthickspace d\theta\,p(\theta|x)\:y(\theta)\Big|.\label{eq:defn-Delta}
\end{equation}
Then, using the same prior, we evaluate the corresponding bias pressure,
$b(\hat{\theta}_{x})$. The figure shows 100 observations $x$ drawn
from $p_{\star}(x)=\int d\theta\,p_{\star}(\theta)\,p(x|\theta)$,
and we believe this justifies the use of the word ``bias'' to describe
$b(\theta)$. The figure is for $d=11$, but a similar relationship
is seen in other dimensionalities.

Instead of looking at particular observations $x$, figure \ref{fig:scores-vs-d}
shows global criteria $I(X;\Theta)$ and $B=\max_{\theta}b(\theta)$.
The optimal prior is largely unaffected by the addition of many irrelevant
dimensions. Once $d>3$, it captures essentially the same information
in any higher dimension, and has zero bias (or near-zero bias, in
our numerical approximation). We may think of this as a new invariance
principle, that predictions should be independent of unobservable
model details. This replaces one of the invariances of Jeffreys, that
repetition of the experiment not change the prior. Repetition-invariance
guarantees poor performance when we are far from the asymptotic limit,
as we see here from the rapidly declining performance of Jeffreys
prior with increasing dimension, capturing less than one bit in $d=26$.
This decline in information is mirrored by a rise in the worst-case
bias $B$.

Figures \ref{fig:posterior-3-26}-\ref{fig:scores-vs-d} also show
a third prior which is log-normal in each decay rate $k_{\mu}=e^{\theta_{\mu}}>0$,
that is, normal in terms of $\theta\in\mathbb{R}^{d}$: 
\begin{equation}
p_{\mathrm{LN}}(\theta)\propto\prod_{\mu=1}^{d}e^{-(\theta_{\mu}-\bar{\theta})^{2}/2\bar{\sigma}^{2}},\qquad\bar{\theta}=0,\:\bar{\sigma}=1\:\forall\mu.\label{eq:log-normal-prior}
\end{equation}
This is not a strongly principled choice, but something like this
is commonly used for parameters known to be positive. Here it produces
better results than Jeffreys prior in high dimensions. We observe
that it also suffers a decline in performance with increasing $d$,
despite making no attempt to deliberately adapt to the high-dimensional
geometry. The details of how well it works will of course depend on
the values chosen for $\bar{\theta},\bar{\sigma}$, and more complicated
priors of this sort can be invented. With enough free ``meta-parameters''
like $\bar{\theta},\bar{\sigma}$, we can surely adjust such a prior
to approximate the optimal prior, and in practice such a variational
approach might be more useful than solving for the optimal prior directly.
We believe that worst-case bias $B=\max_{\theta}b(\theta)$ is a good
score for this purpose, partly because its zero point is meaningful.

\subsection*{Inequivalent parameters}

\noindent Compared to these toy models, more realistic models often
still have many parameter combinations poorly fixed by data, but seldom
come in families which allow us to easily tune the number of dimensions.
Instead of having many interchangeable parameters, each will often
describe a different microscopic effect which we know to exist, even
if we aren't sure which combination of them will matter in a given
regime \citep{hines2014determination}. To illustrate this, we now
examine some models of enzyme kinetics, starting with the famous reaction:
\begin{equation}
E+S\overset{k_{f}}{\underset{k_{r}}{\rightleftharpoons}}ES\overset{k_{p}}{\to}E+P\label{eq:reaction-three-k}
\end{equation}
This summarises differential equations for the concentrations, such
as $\partial_{t}[P]=k_{p}[ES]$ for the final product $P$, and $\partial_{t}[E]=-k_{f}[E][S]+k_{r}[ES]+k_{p}[ES]$
for the enzyme, which combines with the substrate to form a bound
complex.

\begin{figure}
\includegraphics[width=1\columnwidth]{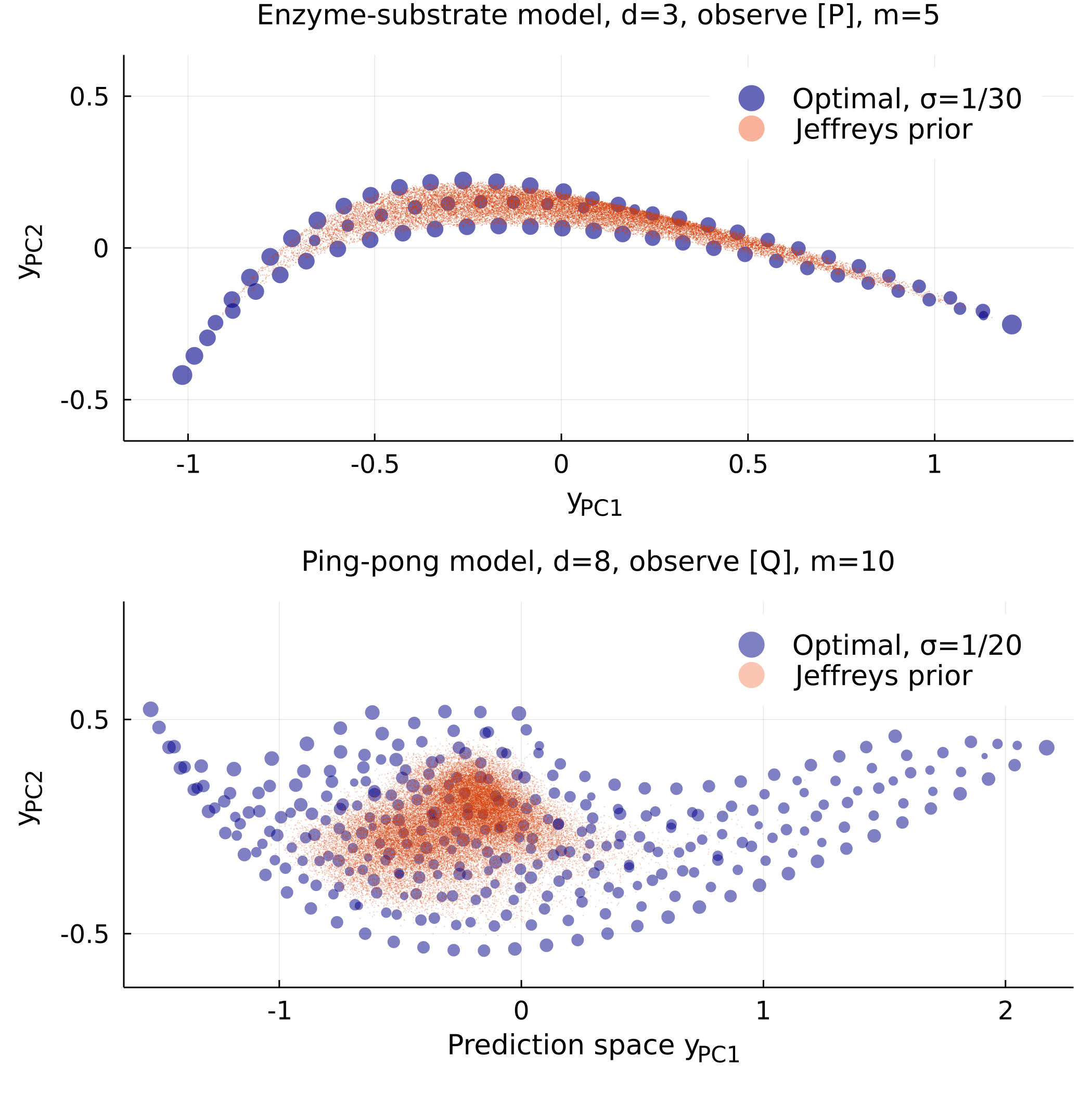}

\caption{Priors for two models of enzyme kinetics. Above, the 3-parameter model
from equation (\ref{eq:reaction-three-k}), observing only the concentration
of product $[P]$ at times $t=1,2,\ldots5$. Below, the 8-parameter
model from equation (\ref{eq:reaction-eight-k}), observing only the
final product $[Q]$ at times $t=1,2,\ldots10$. Here Jeffreys prior
has worst-case bias $B\approx28$ bits, comparable to the models in
figure \ref{fig:scores-vs-d} at similar dimension. While the optimal
prior for the $d=3$ model has its weight on well-known 2-parameter
approximations, including that of Michaelis \& Menten, the edge structure
of the $d=8$ model is much more complicated. (For suitable initial
conditions, it will include the $d=3$ model as an edge.) \label{fig:enzyme-models}}
\end{figure}

If the concentration of product $[P]$ is observed at some number
times, with some noise, and starting from fixed initial conditions,
then this model is not unlike the toy model above. Figure \ref{fig:enzyme-models}
shows the resulting priors for the rate constants appearing in equation
(\ref{eq:reaction-three-k}). The shape of the model manifold is similar,
and the optimal prior again places most of its weight along two one-dimensional
edges, while Jeffreys prior places it in the bulk, favouring the region
where all 3 rate constants come closest to having independently visible
effects on the data. But the resulting bias is not extreme in 3 dimensions.

The edges of this model are known approximations, in which certain
rate constants become infinite (or equal), which we discuss in the
appendix \citep{transtrum2016bridging}. These approximations are
useful in practice since each spans the full length of the most relevant
parameter. But the more difficult situation is when many different
processes of comparable speed are unavoidably involved. The model
manifold may still have many short directions, but the simpler description
selected by $p_{\star}(\theta)$ will tend to have weight on many
different processes. In other words, the simpler model according to
information theory isn't necessarily one simpler model obtained by
taking a limit, but instead, a mixture of many different analytic
limits.

To see this, we consider a slightly more complicated enzyme kinetics
model, the ping-pong mechanism with $d=8$ rate constants:
\begin{align}
E+A\to EA\rightleftharpoons E^{*}P\to E^{*}+P\label{eq:reaction-eight-k}\\
E^{*}+B\to E^{*}B & \rightleftharpoons EQ\to E+Q.\nonumber 
\end{align}
Here $E^{*}$ is a deformed version of the enzyme $E$, which is produced
in the reaction from $A$ to $P$, and reverted in the reaction from
$B$ to final product $Q$. There are clearly many more possible limits
in which some combination of the rate constants become large or small.
Figure \ref{fig:enzyme-models} shows that the optimal prior has weight
on at least five different 1-edges, none of which is a good description
by itself.

The concentration of weight seen in Jeffreys prior for these enzyme
models is comparable what we had before, with worst-case bias pressure
$B\approx14$ bits in $d=3$ and $28$ bits in $d=8$. These examples
share geometric features with many real models in science \citep{machta2013parameter},
and thus we believe the problems described here are generic.

\section*{Discussion}

\noindent Before fitting a model to data there is often a selection
step, to choose a model which is complex enough to fit the true pattern,
but not so complex as to fit the noise. The motivation for this is
clear in maximum likelihood estimation, where only one $\hat{\theta}_{x}$
is kept, and there are various criteria for making the trade-off \citep{Akaike:1974ih,rissanen1978modeling,grunwald2019minimum}.
The motivation is less clear in Bayesian analysis, where slightly
different criteria can be derived by approximating $p(x)$ \citep{Schwarz:1978uv,myung2000counting}.
We might hope that if many different points $\theta$ are consistent
with the noisy data $x$, then the posterior $p(\theta|x)$ should
simply have weight on all of them, encoding our uncertainty about
$\theta$.

Why then is model selection needed at all in Bayesian inference? Our
answer here is that this is done to avoid measure-induced bias, not
overfitting. When using a sub-optimal prior, models with too much
complexity do indeed perform badly. This problem is seen in figure
\ref{fig:scores-vs-d}, in the rapid decline of scores $I(X;\Theta)$
or $B$ with increasing $d$, and would also be seen in the more traditional
model evidence $p(x)$ --- all of these scores prefer models with
$d\leq3$. But the problem is not overfitting, since the extra parameters
being added are irrelevant, i.e. they can have very little effect
on the predictions $y_{t}(\theta)$. Instead, the problem is concentration
of measure. In models with tens of parameters this effect can be enormous:
It leads to posterior expectation values $\Delta>20$ standard deviations
away from ideal, for the $d=26$ model with Jeffreys prior, and mutual
information $I<1$~bit learned, and $B>500$~bits of bias. This
problem is completely avoided by the optimal prior $p_{\star}(\theta)$,
which suffers no decline in performance with increasing parameter
count $d$.

Geometrically, we can view traditional model selection as adjusting
$d$ to ensure that the model manifold only has dimensions of length
$L>1$. This ensures that most of the posterior weight is in the interior
of the manifold, hence ignoring model edges is justified. By contrast,
when there are dimensions of length $L<1$, the optimal posterior
will usually have its weight at their extreme values --- on several
manifold edges, which are themselves simpler models \citep{mattingly2018maximizing}.
Fisher lengths $L$ depend on the quantity of data to be gathered,
and repeating an experiment $M$ times enlarges all by a factor $\sqrt{M}$.
Large enough $M$ can eventually make any dimension larger than 1,
and thus repetition alters what $d$ traditional model selection prefers.
Similarly, repetition alters the effective dimensionality of $p_{\star}(\theta)$.
Some earlier work on model geometry studies a series in $1/M$ \citep{balasubramanian1997statistical,myung2000counting,piasini2022effect};
this expansion around $L=\infty$ captures some features beyond the
volume but is not suitable for models with dimensions $L\ll1$.

Real models in science typically have many irrelevant parameters \citep{machta2013parameter,oleary2013correlations,wen2017forcematching,marschmann2019equifinality,karakida2021pathological}.
It is common to have parameter directions $10^{-10}$ times as important
as the most relevant one, but impossible to repeat an experiment the
$M=10^{20}$ times needed to bridge this gap. Sometimes it is possible
to remove the irrelevant parameters, and derive a simpler effective
theory. This is what happens in physics, where a large separation
of scales allows great simplicity and high accuracy \citep{Kadanoff:1966wm,Wilson:1971bg}.
But many other systems we would like to model cannot, or cannot yet,
be so simplified. For complicated biological reactions, or climate
models, or neural networks, it is unclear which of the microscopic
details can be safely ignored, or what the right effective variable
are. Unlike our toy models, we cannot easily adjust $d$, since every
parameter has a different meaning. This is why we seek statistical
methods which do not require us to find the right effective theory.
And in particular, here we study priors almost invariant to complexity.

The optimal prior is discrete, which makes it difficult to find, and
this difficulty appears to be why its good properties have been overlooked.
It is known analytically only for extremely simple models like $M=1$
Bernoulli, and previous numerical work only treated slightly more
complicated models, with $d\leq2$ parameters \citep{mattingly2018maximizing}.
While our concern here is with the ideal properties, for practical
use nearly-optimal approximations may be required. One possibility
is the adaptive slab-and-spike prior introduced in \citep{quinn2023information}.
Another would be to use some variational family $p_{\lambda}(\theta)$
with adjustable meta-parameters $\lambda$ \citep{nalisnick2017learning}.

Discreteness is also how the exactly optimal $p_{\star}(\theta)$
encodes a length scale $L\approx1$ in the model geometry, which is
the divide between relevant and irrelevant parameters, between parameters
which are constrained by data and those which are not. Making this
distinction in some way is essential for good behaviour, and it implies
a dependence on the quantity of data. An effective model appropriate
for much less data than observed will be too simple: The atoms of
$p_{\mathrm{\star}}(\theta)$ will be too far apart (much like recording
too few significant figures), or else selecting a small $d$ means
picking just one edge (fixing some parameters which may in fact be
relevant). On the other hand, what we have demonstrated here is that
a model appropriate for much more data --- infinitely much in the
case of $p_{\mathrm{J}}(\theta)$ --- will instead introduce enormous
bias into our inference about $\theta$. 

\section*{Acknowledgements}

\noindent We thank Isabella Graf, Mason Rouches, Jim Sethna and Mark
Transtrum for helpful comments on a draft.

Our work is supported by a Simons Investigator Award. Parts of this
work were performed at the Aspen Center for Physics, supported by
NSF grant PHY-1607611. The participation of MCA at Aspen was supported
by the Simons Foundation.

\section*{Appendix}

\noindent For brevity the main text omits some standard definitions.
The KL divergence (or relative entropy) is defined
\[
D_{\mathrm{KL}}\big[p(x)\big\Vert q(x)\big]=\int dx\:p(x)\,\log\frac{p(x)}{q(x)}.
\]
With a conditional probability, in our notation $D_{\mathrm{KL}}\big[p(x|\theta)\big\Vert q(x)\big]$
integrates $x$ but remains a function of $\theta$. The Fisher information
metric $g_{\mu\nu}(\theta)$ is the quadratic term from expanding
$D_{\mathrm{KL}}\big[p(x|\theta+d\theta)\big\Vert p(x|\theta)\big]$.

The mutual information is 
\begin{align*}
I(X;\Theta) & =D_{\mathrm{KL}}\big[p(x,\theta)\big\Vert p(x)p(\theta)\big]\\
 & =S(\Theta)-S(\Theta|X)\\
 & =\iint dx\:d\theta\:p(x|\theta)p(\theta)\log\frac{p(x|\theta)}{p(x)}
\end{align*}
where we use Bayes theorem,

\[
p(\theta|x)=p(x|\theta)p(\theta)\big/p(x)
\]
with $p(x)=\int d\theta\,p(x|\theta)p(\theta)$, and entropy
\[
S(X)=-\int dx\,p(x)\log p(x).
\]
Conditional entropy is $S(X|\theta)=-\int dx\,p(x|\theta)\log p(x|\theta)$
for one value $\theta$, or $S(X|\Theta)=\int d\theta\,p(\theta)\,S(X|\theta)$.
With a Gaussian likelihood, equation (\ref{eq:defn-gauss-like}),
and $X=\mathbb{R}^{m}$, this is a constant:
\[
S(X|\Theta)=\frac{m}{2}(1+\log2\pi\sigma^{2}).
\]

Many of these quantities depend on the choice of prior, such as the
posterior $p(\theta|x)$ and the mutual information $I(X;\Theta)$.
This is also true of our bias pressure $b(\theta)$ and worst-case
$B=\max_{\theta}b(\theta)$. When we need to refer to those for a
specific prior such as $p_{\star}(\theta)$, we use the same subscript,
writing $I_{\mathrm{\star}}$ and $B_{\mathrm{\star}}.$

All probability distributions are normalised. In particular, the gradient
in (\ref{eq:defn-bias}) is taken with the constraint of normalisation
--- varying the density at each point independently would give a
different constant.

\subsection{Square hypercone}

\noindent Here we consider an even simpler toy model, in which we
can more rigorously define what we mean by co-volume, and analytically
calculate the posterior deviation.

Consider a $d$-dimensional cone, consisting of a line of length $L$
thickened to have a square cross section. This is $y(\theta)=(\theta_{1},r\theta_{2},r\theta_{3},\ldots)$
with scale $r(\theta_{1})=\theta_{1}/L$, and co-ordinate ranges:
\begin{align*}
0 & \leq\theta_{1}\leq L\\
0 & \leq\theta_{\mu}\leq1,\quad\mu=2,3,\ldots,d.
\end{align*}
Fixing noise $\sigma=1$, this has one relevant dimension, length
$L$, and $d-1$ irrelevant dimensions whose lengths are always $\leq1$.
The FIM is then: 
\[
g(\theta)=\left[\begin{array}{cccc}
1+\sum_{\mu=2}^{d}r^{2} & \frac{\theta_{1}\theta_{2}}{L^{2}} & \frac{\theta_{1}\theta_{3}}{L^{2}} & \cdots\\
\frac{\theta_{1}\theta_{2}}{L^{2}} & r^{2} & 0\\
\frac{\theta_{1}\theta_{3}}{L^{2}} & 0 & r^{2}\\
\vdots &  &  & \ddots
\end{array}\right].
\]
Thus the volume element is 
\[
\sqrt{\det g(\theta)}=1\:r(\theta_{1})^{(d-1)}.
\]
Regarding the first factor as $\sqrt{\det g_{\mathrm{rel}}}+\mathcal{O}(1/L^{2})$,
it is trivial to integrate the second factor over $\text{\ensuremath{\theta_{\mu}}}$
for all $\mu\geq2$, and this factor $\int_{0}^{1}d\theta_{2}\cdots\int_{0}^{1}d\theta_{d}\sqrt{\det g_{\mathrm{irrel}}}=(r(\theta_{1}))^{(d-1)}$
is the irrelevant co-volume. The effective Jeffreys prior along the
one relevant dimension is thus 
\[
p_{\mathrm{J}}(\theta_{1})\propto(\theta_{1})^{(d-1)}
\]
which clearly has much more weight at large $\theta_{1}$, at the
thick end of the cone.

Now observe some $x$, giving $p(\theta_{1}|x)\propto e^{-(x-\theta_{1})^{2}/2}(\theta_{1})^{(d-1)}$.
With a few lines of algebra we can derive, assuming $1\ll x\ll L$,
that the posterior deviation (\ref{eq:defn-Delta}) is 
\[
\Delta=x-\left\langle \theta_{1}\right\rangle _{p(\theta_{1}|x)}=\frac{d-1}{x}+\mathcal{O}\Big(\frac{1}{x^{3}}\Big).
\]

Choosing $L=50$ and $d=26$ to roughly match figure \ref{fig:posterior-3-26},
at $x\approx10$ the deviation is $\Delta\approx2.5$$\Delta\approx1.25$.
This is smaller than what is seen for the exponential decay model,
whose geometry is of course more complicated. This difference is also
detected by bias pressure. The maximum $b(\theta)$ for this cone
is about 55 bits, which is close to the $d=5$ model in figure \ref{fig:scores-vs-d}.

\begin{figure}
\includegraphics[width=1\columnwidth]{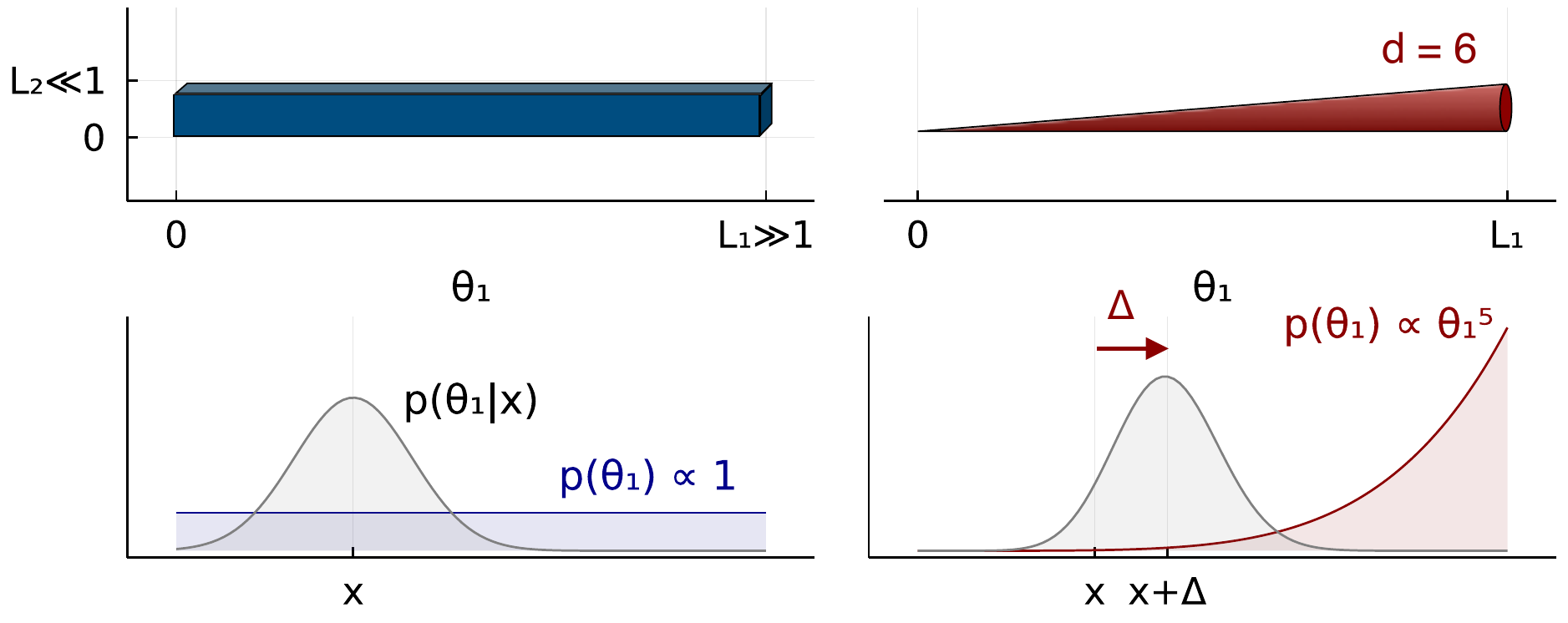}

\caption{The simplest geometry in which to see measure-induced posterior bias.
We compare two $d=6$ model manifolds, both with $\theta_{1}$ relevant,
Fisher length $L_{1}\gg1$, and five irrelevant dimensions ($L_{\mu}\ll1$,
$\mu=2,3,\ldots6$) which are either of constant size (left) or taper
linearly (right). The distinction between these two situations is
by assumption unobservable, but using the $d$-dimensional notion
of volume as a prior gives an effective $p(\theta_{1})$ which is
either flat, or $\propto(\theta_{1})^{5}$. This can induce substantial
bias in the posterior $p(\theta_{1}|x)$. \label{fig:cone-cube}}
\end{figure}

While this example takes all irrelevant dimensions to be of equal
Fisher length, it would be more realistic to have a series $L,1,L^{-1},L^{-2},\ldots$
equally spaced on a log scale. This makes no difference to the effective
$p_{\mathrm{J}}(\theta_{1})\propto(\theta_{1})^{(d-1)}$ and hence
to the posterior deviation $\Delta$. Figure \ref{fig:cone-cube}
draws instead a cone with round cross-section, which also makes no
difference. It compares this to a shape of constant cross-section,
for which $p_{\mathrm{J}}(\theta_{1})\propto1$, hence there is no
such bias.

\subsection{Estimating $I(X;\Theta)$ and its gradient\label{subsec:Estimating-MI}}

\noindent The mutual information can be estimated, for a discrete
prior 
\[
p(\theta)=\sum_{a=1}^{K}\lambda_{a}\delta(\theta-\theta_{a})
\]
and a Gaussian likelihood (\ref{eq:defn-gauss-like}), by replacing
the integral over $x$ with normally distributed samples:
\[
I(X;\Theta)=\sum_{a}\lambda_{a}\negthickspace\negthickspace\underset{\text{sample }x\sim\mathcal{N}(\theta_{a},\sigma^{2})}{\underbrace{\int dx\:p(x|\theta_{a})}}\negthickspace\negthickspace\log\frac{p(x|\theta_{a})}{\sum_{b}\lambda_{b}p(x|\theta_{b})}.
\]
This gives an unbiased estimate, and what we used for plotting $I(X;\Theta)$
in figures \ref{fig:vary-sigma} and \ref{fig:scores-vs-d}. 

But for finding $p_{\star}(\theta)$, what we need is the very small
gradients of $I$ with respect to each $\theta_{a}$: near to the
optimum, the function is very close to flat. We find that, instead
of Monte Carlo, the following kernel density approximation works well:
\begin{align}
S(X) & =-\sum_{a}\lambda_{a}\int dx'\:p(x'|\theta_{a})\log\sum_{b}\lambda_{b}\:p(x'|\theta_{b})\nonumber \\
 & =-\sum_{a}\lambda_{a}\int dx'\:p(x'|\theta_{a})\Big[\log\sum_{b}\lambda_{b}\:p(x=\theta_{a}|\theta_{b})\nonumber \\
 & \qquad\qquad\qquad\qquad+\mathcal{O}(x'-\theta_{a})^{2}\Big]\nonumber \\
 & \approx-\sum_{a}\lambda_{a}\log\sum_{b}\lambda_{b}\:e^{[y(\theta_{a})-y(\theta_{b})]^{2}/2\sigma^{\prime2}}+\text{const.}\label{eq:S-KDE}
\end{align}
Here we Taylor expand the $\log p(x)$ about $x=\theta_{a}$ for each
atom \citep{Huber:2008ju}. For the purpose of finding $p_{\star}(\theta)$
we may ignore the constant, and the conditional entropy in $I(X;\Theta)=S(X)-S(X|\Theta)$.
Further, this is a better estimate used at $\sigma'=\sqrt{2}\sigma$.

Before maximising $I(X;\Theta)$ using L-BFGS \citep{Nocedal:1980hv,johnsonnlopt}
to adjust all $\theta_{a}$ and $\lambda_{a}$ together, we find it
useful to sample initial points using Mitchell's best-candidate algorithm
\citep{mitchell1991spectrally}.

\subsection{Other methods}

\noindent Our focus here is on the properties of the optimal prior
$p_{\star}(\theta)$, but better ways to find nearly optimal solutions
may be needed in order to use these ideas on larger problems. Some
ideas have been explored in the literature:
\begin{itemize}
\item The famous algorithm for finding $p_{\star}(\theta)$ is due to Blahut
\& Arimoto \citep{Blahut:1972ed,Arimoto:1972jz}, but it needs a discrete
$\Theta$ which limits $d$. This was adapted to use MCMC sampling
instead by \citep{lafferty2001iterative}, although their work appears
to need discrete $X$ instead. Perhaps it can be generalised.
\item More recently, the following lower bound for $I(X;\Theta)$ was used
by \citep{nalisnick2017learning} to find approximations to $p_{\star}(\theta)$:
\[
I(X;\Theta)\geq I_{\mathrm{NS}}=-S(X|\Theta)-\int\negmedspace dx\:p(x)\log\max_{\theta}p(x|\theta).
\]
 This bound is too crude to see the features of interest here: For
all models in this paper, it favours a prior $p_{2}(\theta)=\argmax_{p(\theta)}I_{\mathrm{NS}}$
with just two delta functions, for any noise level $\sigma$.
\item We mentioned above that adjusting some ``meta-parameters'' of some
distribution $p_{\bar{\theta},\bar{\sigma}}(\theta)$ would be one
way to handle near-optimal priors. This is the approach of \citep{nalisnick2017learning},
and of many papers maximising other scores, often described as ``variational''.
\item Another prior which typically has large $I(X;\Theta)$ was introduced
in \citep{quinn2023information} under the name ``adaptive slab-and-spike
prior''. It pulls every point $x$ in a distribution 
\[
p_{\mathrm{NML}}(x)=\frac{\max_{\hat{\theta}}p(x|\hat{\theta})}{Z},\quad Z=\int\negmedspace dx\:{\textstyle \max_{\hat{\theta}}p(x|\hat{\theta})}
\]
 back to its maximum likelihood point $\hat{\theta}$:
\[
p_{\mathrm{proj}}(\theta)=\int\negmedspace dx\:p_{\mathrm{NML}}(x)\:\delta\big(\theta-\argmax_{\hat{\theta}}p(x|\hat{\theta})\big).
\]
The result has weight everywhere in the model manifold, but extra
weight on the edges. Because the amount of weight on edges is controlled
by $\sigma$, it adopts an appropriate effective dimensionality (\ref{eq:d_eff-scaling}),
and has low bias (\ref{eq:defn-bias}).
\end{itemize}

\subsection{Bias alla Bennett\label{subsec:Jarzynski}}

\noindent The KL divergence integral needed for $b(\theta)$ is somewhat
badly behaved when the prior's weight is far from the point $\theta$.
To describe how we handle this, we begin with the na\"ive Monte Carlo
calculation of $p(x)=\int d\theta\,p(\theta)\,p(x|\theta)$, which
involves sampling from the prior. If $x$ is very far from where the
prior has most of its weight, then we will never get any samples where
$p(x|\theta)$ is not exponentially small, so we will miss the leading
contribution to $p(x)$.

Sampling from the posterior $p(\theta|x)\propto p(\theta)\,p(x|\theta)$
instead, we will get points in the right area, but the wrong answer.
The following identity due to Bennett \citep{bennett1976efficient}
lets us walk from the prior to the posterior and get the right answer,
sampling from distributions $\propto p(\theta)\,p(x|\theta)^{\alpha}$
for several powers $\alpha$. Defining $\Delta_{\alpha}^{\delta}(x)$:
\begin{align*}
0 & =\alpha_{0}<\alpha_{1}<\ldots<\alpha_{n}=1\\
e^{\Delta_{\alpha}^{\delta}(x)} & =\left\langle p(x|\theta)^{\delta}\right\rangle _{x,\alpha}=\frac{\int d\theta\,p(\theta)\,p(x|\theta)^{\alpha}\:p(x|\theta)^{\delta}}{\int d\theta\,p(\theta)\,p(x|\theta)^{\alpha}}
\end{align*}
the result is 
\[
\log p(x)=\sum_{i=0}^{n-1}\Delta_{\alpha_{i}}^{(\alpha_{i+1}-\alpha_{i})}(x)=-\sum_{i=1}^{n}\Delta_{\alpha_{i}}^{(\alpha_{i-1}-\alpha_{i})}(x)
\]
i.e.
\[
p(x)=\prod_{\alpha\neq1}\left\langle p(x|\theta)^{\alpha_{\mathrm{next}}-\alpha\vphantom{1}}\right\rangle _{x,\alpha}=1\Big/\prod_{\alpha\neq0}\left\langle 1\big/p(x|\theta)^{\alpha-\alpha_{\mathrm{prev}}\vphantom{1}}\right\rangle _{x,\alpha}.
\]
For $n=1$ the first is trivial, and the second reads $1/p(x)=\left\langle 1/p(x|\theta)\right\rangle _{\theta\sim p(\theta|x)}$.

Next, we want $D_{\mathrm{KL}}[p(x|\varphi)\Vert p(x)]$ which involves
$-\int dx\,p(x|\varphi)\,\log p(x)$. To plug this in, we would need
to average $\left\langle \ldots\right\rangle _{x,\alpha}$ at every
$x$ in the integral. Rather than sample from a fresh set of $\alpha$-distributions
for each $x$, we can take one set at some $x_{0}$, and correct using
importance sampling to write:
\[
\left\langle \ldots\right\rangle _{x,\alpha}=\left\langle \ldots\frac{p(x|\theta)}{p(x_{0}|\theta)}\right\rangle _{x_{0},\alpha}
\]
All of this can be done without knowing the normalisation of the prior,
$Z$ in (\ref{eq:defn-Jeff}).

The same set of samples from $\alpha$-distributions can be used to
calculate either forwards or backwards. These give upper and lower
bounds on the true value. When they are sufficiently far apart to
be visible, the plots show both of them. Figure \ref{fig:vary-sigma}
uses, as the $\alpha$-distributions, all larger $\sigma$-values
on the plot, and also shows (dotted line) the result of na\"ive sampling
from $p_{\mathrm{J}}(\theta)$. Figure \ref{fig:scores-vs-d} uses
about 30 steps.

\subsection{Jeffreys \& Vandermonde\label{subsec:Vandermonde}}

\noindent To find Jeffreys prior, let us parameterise the exponential
decay model (\ref{eq:defn-exp-model}) by $\phi_{\mu}=e^{-\exp(\theta_{\mu})}$
which lives in the unit interval:
\[
y_{t}(\theta)=\sum_{\mu=1}^{d}\frac{e^{-k_{\mu}t}}{d}=\sum_{\mu=1}^{d}\frac{(\phi_{\mu})^{t}}{d}
\]
where $\theta_{\mu}=\log k_{\mu}\in\mathbb{R},\;0\leq\phi_{\mu}\leq1$.
The Fisher information metric (\ref{eq:defn-FIM}) reads $g_{\mu\nu}=\frac{1}{\sigma^{2}}\sum_{t}^{m}J_{\mu t}J_{\nu t}$
in terms of a Jacobian which, in these co-ordinates, takes the simple
form:
\[
J_{\mu t}(\phi)=\frac{\partial y_{t}}{\partial\phi_{\mu}}=\frac{1}{d}t\:\phi^{t-1}.
\]
In high dimensionality, $g_{\mu\nu}$ is very badly conditioned, and
thus it is difficult to find the determinant with sufficient numerical
accuracy (or at least, it is slow, as high-precision numbers are required).
However, in the case $d=m$ where the Jacobian is a square matrix,
it is of Vandermonde form. Hence its determinant is known exactly,
and we can simply write:
\begin{align*}
p_{\mathrm{J}}(\phi) & =\sqrt{\det g_{\mu\nu}(\phi)}=\left|\det J_{\mu t}(\phi)\right|\\
 & =\prod_{1\leq\mu<\nu\leq d}\left|\phi_{\mu}-\phi_{\nu}\right|\prod_{t=1}^{m}\frac{t}{d}.
\end{align*}
For $d\neq m$, more complicated formulae (involving a sum over Schur
polynomials) are known, but the points in figure \ref{fig:scores-vs-d}
are chosen not to need them.

To sample from Jeffreys prior, or posterior, we use the affine-invariant
``emcee'' sampler \citep{Goodman:2010et}. This adapts well to badly
conditioned geometries. Because the Vandermonde formula lets us work
in machine precision, we can sample $10^{6}$ points in a few minutes,
which is sufficient for figure \ref{fig:posterior-3-26}.

In the enzyme kinetics models of figure \ref{fig:enzyme-models},
finding $y_{t}(\theta)$ involves solving a differential equation.
The gradient $\partial y_{t}/\partial\theta_{\mu}$ is needed both
for Jeffreys density, and for maximising $S(X)$ via the above KDE
formula, (\ref{eq:S-KDE}). This can be handled efficiently by passing
dual numbers through the solver \citep{ma2021comparison}.

\subsection{Michaelis--Menten et. al.\label{subsec:Michaelis=002013Menten}}

\noindent The arrows in (\ref{eq:reaction-three-k}) summarise the
following differential equations for the concentrations of the four
chemicals involved:
\begin{align*}
\partial_{t}[E] & =-k_{f}[E][S]+k_{r}[ES]+k_{p}[ES]\\
\partial_{t}[S] & =-k_{f}[E][S]+k_{r}[ES]\\
\partial_{t}[ES] & =+k_{f}[E][S]-k_{r}[ES]-k_{p}[ES]\\
\partial_{t}[P] & =k_{p}[ES].
\end{align*}
These equations conserve $E_{0}=[E]+[ES]$ (the enzyme is recycled)
and $S_{0}=[S]+[ES]+[P]$ (the substrate is converted to product)
leaving two dynamical quantities, $[S]$ and $[P]$. The plot takes
them to have initial values $[S]=1$, $[P]=0$ at $t=0$, and we fix
$E_{0}=1/4$, $S_{0}=1$. 

The original analysis of Michaelis \& Menten \citep{Menten:1913wn}
takes the first two reactions to be in equilibrium. This can be viewed
as taking the limit $k_{r},k_{f}\to\infty$ holding fixed $K_{D}=k_{r}/k_{f}$,
which picks a 2-parameter subspace of $\Theta$, an edge of the manifold.
Then $[ES]=[E][S]/K_{D}$ becomes constant, leaving their equation
\begin{equation}
\partial_{t}[P]=\frac{k_{p}E_{0}[S]}{K_{D}+[S]}.\label{eq:MM-equilib}
\end{equation}
If we do not observe $[E]$, then this is almost identical to the
quasi-static limit of Briggs \& Haldane \citep{briggs1925note} who
take $k_{r},E_{0}\to0$, and $k_{f},k_{p}\to\infty$ holding fixed
$K_{M}=k_{p}/f_{f}$ and $V_{\mathrm{max}}=k_{p}E_{0}$, which gives
\[
\partial_{t}[P]=\frac{V_{\mathrm{max}}[S]}{K_{M}+[S]}
\]
which was much later shown to be analytically tractable \citep{Schnell:1997gn}.

In figure \ref{fig:enzyme-models}, most of the points of weight in
the optimal prior lie on the intersection of these two 2-parameter
models, that is, on a pair of one-parameter models. These and other
limits were discussed geometrically in \citep{transtrum2016bridging}.

For the $d=8$ ping-pong model (\ref{eq:reaction-eight-k}), we take
initial conditions of $[A]=[B]=1$, $[E]=[E*]=0.5$, $[EA]=[E*P]=[E*B]=[EQ]=0.1$,
$[P]=[Q]=0$.

\subsection{Ever since the big bang}

\noindent The claim that the age of the universe constrains us from
taking the asymptotic limit in realistic multi-parameter models deserves
a brief calculation. The conventional age is $13.8\times10^{8}$ years,
about $4.4\times10^{17}$ seconds \citep{planckcollaboration2021planck}.

For the model (\ref{eq:defn-exp-model}), manifold widths are shown
to scale like $L\propto\sqrt{\mathop{\mathrm{eig}}g}$ in \citep{transtrum2010why},
and \citep{transtrum2010why,machta2013parameter} shows FIM eigenvalues
$\propto10^{-d}$. Repeating an experiment $M$ times scales lengths
by $\sqrt{M}$, so we need roughly $10^{d}$ repetitions to make the
smallest manifold width larger than 1. If the initial experiment took
1 second, then it is impossible to perform enough repetitions to make
all dimensions of the $d=26$ model relevant.

Other models studied in \citep{transtrum2010why,machta2013parameter}
have different slopes $\log(\mathop{\mathrm{eig}}g)$ vs. $d$, so
the precise cutoff will vary. But models with hundreds, or thousands,
of parameters are also routine. It seems safe to claim that for most
of these, the asymptotic limit is never justified.

\subsection{Terminology}

\noindent What we call Jeffreys prior, $p_{\mathrm{J}}(\theta)$,
is (apart from variations in apostrophes and articles) sometimes called
``Jeffreys's nonlocation rule'' \citep{kass1996selection} in order
to stress that it is not quite what Jeffreys favoured. He argued for
separating ``location'' parameters (such as an unconstrained position
$\phi\in\mathbb{R}$) and excluding them from the determinant. The
models we consider here have no such perfect symmetries.

What we call the optimal prior $p_{\star}(\theta)$ is sometimes called
``Shannon optimal'' \citep{scholl1998shannon}, and sometimes called
a ``reference prior'' after Bernardo \citep{bernardo1979reference}.
The latter is misleading, as definition 1 in \citep{bernardo1979reference}
explicitly takes the asymptotic limit $M\to\infty$. Which is then
Jeffreys prior, until some ways to handle ``nuisance parameters''
are appended. The idea of considering $\argmax_{p(\theta)}I(X;\Theta)$
in the limit is older, for instance Lindley \citep{lindley1961use}
considers it, and also notes that it leads to Jeffreys prior (which
he does not like for multi-parameter models, but not for our reasons).
In \citep{berger1989priors} the discrete prior for $k$ repetitions
is called the ``$k$-th reference prior'', and is understood to
be discrete, but is of interest only as a tool for showing that the
limit exists. We stress that the asymptotic limit removes what are
(for this paper) the interesting features of this prior.

The same $p_{\star}(\theta)$ can also be obtained by an equivalent
minimax game, for which Kashyap uses the term ``optimal prior''
\citep{kashyap1971prior}. But he, too, takes the asymptotic limit.

\subsection{Opened source}

\noindent The code used to find the priors (and the scores) shown
is available at \url{https://github.com/mcabbott/AtomicPriors.jl}.

\bibliographystyle{/Users/me/Documents/Current-Work/0000-zotero-export/my-JHEP-4.bst}
\bibliography{/Users/me/Documents/Current-Work/0000-zotero-export/My-Library,/Users/me/Dropbox/Library.papers3/complete3}

\end{document}